\documentclass[a4paper]{jpconf}
\usepackage{graphicx}
\usepackage{amsmath,amssymb}
\newcommand{\MeV}{\ensuremath{\mathrm{MeV}}}
\newcommand{\GeV}{\ensuremath{\mathrm{GeV}}}

\begin{document}
\title{Dileptons in a coarse-grained transport approach}
\author{ Hendrik van Hees$^{1,2}$, Stephan Endres$^{1,2}$, Janus Weil$^{2}$ and Marcus Bleicher$^{1,2}$}
\address{$^{1}$Institut f\"{u}r Theoretische Physik, Universit\"{a}t
  Frankfurt, Max-von-Laue-Stra{\ss}e 1, 60438 Frankfurt, Germany} 
\address{$^{2}$Frankfurt Institute for Advanced Studies,
  Ruth-Moufang-Stra{\ss}e 1, 60438 Frankfurt, Germany} 
\ead{endres@th.physik.uni-frankfurt.de}
\begin{abstract}
  We calculate dilepton spectra in heavy-ion collisions using a
  coarse-graining approach to the simulation of the created medium with
  the UrQMD transport model. This enables the use of dilepton-production
  rates evaluated in equilibrium quantum-field theory at finite
  temperatures and chemical potentials.  
\end{abstract}

Due to their penetrating nature, the invariant-mass and
transverse-momentum spectra of dileptons are considered unique probes
for the in-medium properties of hadrons in strongly interacting hot and
dense matter, created in heavy-ion collisions
\cite{rw99,Rapp:2009yu}. The dilepton-production rates are proportional
to the electromagnetic current-correlation function of the source
\cite{MT84} and thus provide insight into the corresponding excitations
carrying the quantum numbers of the electromagnetic
current. Particularly from the restoration of chiral symmetry one
expects substantial modifications of the spectral shape of this
current-correlation function. Particularly one hopes to gain insight
into the microscopic realization of the chiral-symmetry restoration,
e.g., whether it is due to dropping in-medium hadron masses as predicted
by the Brown-Rho-scaling conjecture or models implementing the vector
manifestation of chiral symmetry or substantial broadening of the
pertinent light-vector-meson spectral shapes due to hadronic
interactions within the medium (``melting-resonance scenario'') as
predicted by quantum-field-theoretical effective hadronic many-body
theory (for a review see \cite{rw99}).

Since the dileptons are produced and emitted undisturbed by final-state
interactions during the entire evolution of the hot and dense medium,
first a good understanding of all relevant microscopic sources of
electromagnetic radiation in this medium is important. 

For heavy-ion collisions at ultrarelativistic energies, in the early hot
stage the relevant degrees of the medium are quarks and gluons forming a
quark-gluon plasma (QGP) close to local thermal equilibrium, as is
inferred from the successful description of the hadronic bulk
observables like $p_T$ spectra and angular anisotropies $v_n$ by
hydrodynamical models for the evolving medium. Here the main source of
dileptons is the quark-antiquark annihilation
$q+\overline{q} \rightarrow \ell^+ + \ell^-$ (with $\ell=\e,\mu$). In
the later stages of the fireball evolution (or, in the case of heavy-ion
collisions at lower energies, the entire evolution) the medium is
described as a hot hadronic gas. In the low invariant-mass range,
$2m_{\ell} \leq M \lesssim 1\GeV$ the dilepton sources are thus hadronic
and phenomenologically well described within a vector-meson dominance
(VMD) model \cite{sak60}, i.e., the assumption that the electromagnetic
current of the hadrons is proportional to the light-vector meson
field. This also implies that the electromagnetic transition form
factors for the various Dalitz decays of hadron resonances contributing
substantially to the dilepton yield in the low-mass tail are describable
within the VMD model. In the intermediate-mass range also ``multi-pion
processes'' become relevant.
\begin{figure}
\begin{minipage}{0.48\linewidth}
\includegraphics[width=\linewidth]{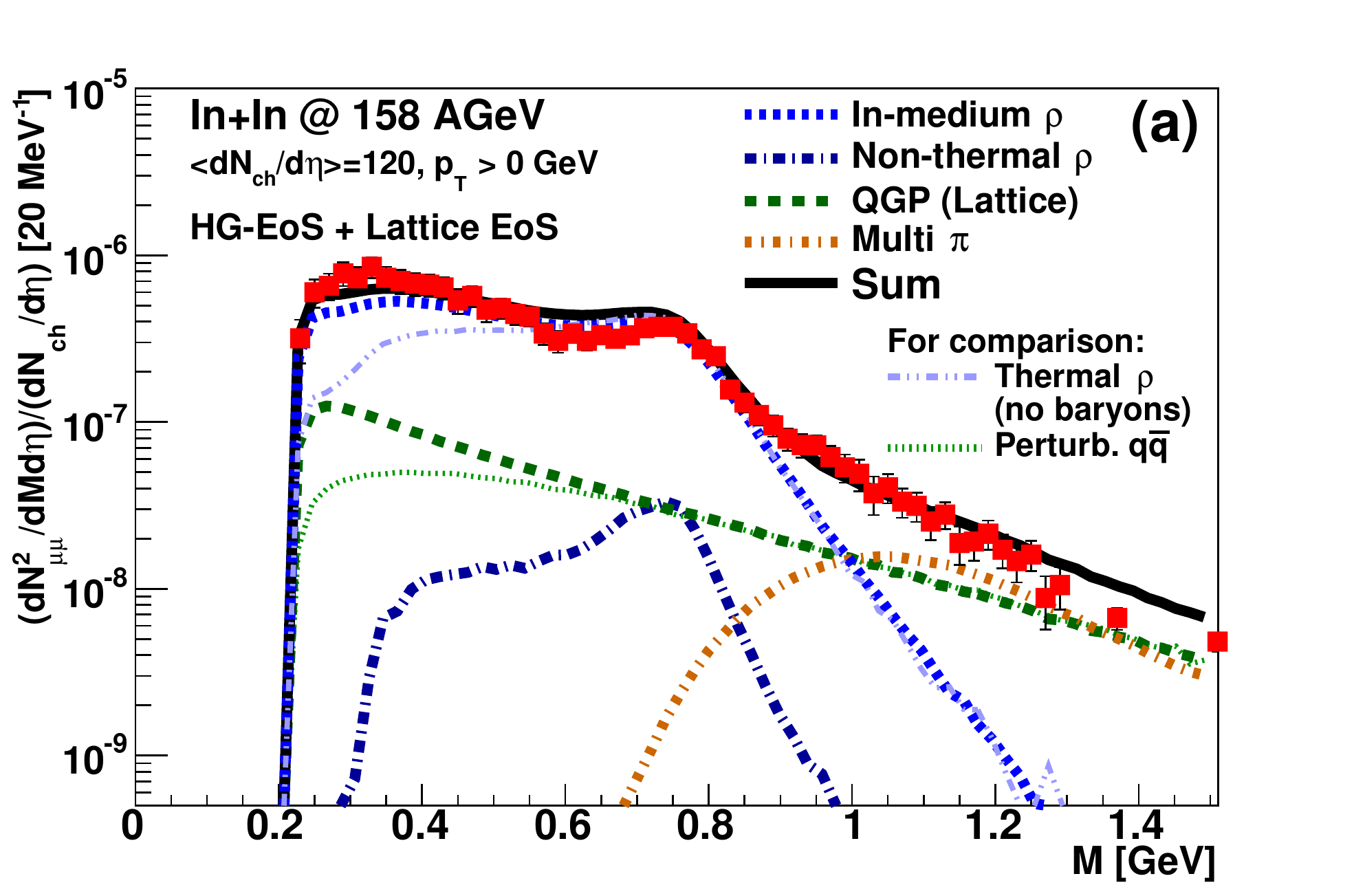}
\end{minipage} \hfill
\begin{minipage}{0.48\linewidth}
\includegraphics[width=1.02\columnwidth]{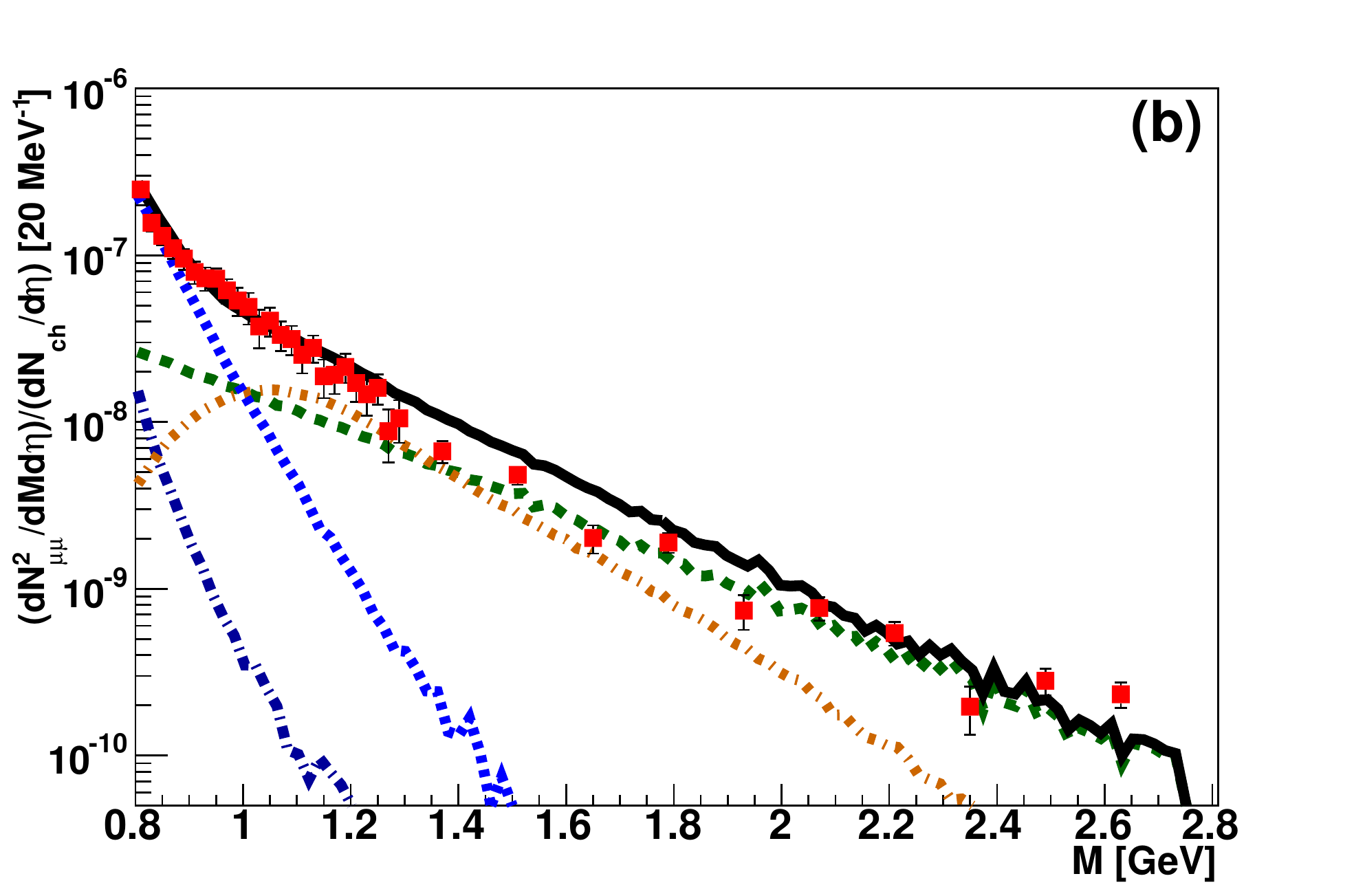}
\end{minipage}
\caption{The dimuon spectra in $158A \, \GeV$ In+In collisions at the
  CERN SPS in (a) the low-mass (LMR) and (b) the intermediate-mass
  region (IMR) in comparison to the data from the NA60 Collaboration
  \cite{Arnaldi:2008er, Specht:2010xu, Arnaldi:2008fw}. In the LMR the
  dilepton yield is dominated by the decays of the in-medium $\rho$
  mesons (dotted blue line) and in the IMR by QGP and multi-pion
  hadronic processes. The comparison with a dilepton rate based on a
  thermal-$\rho$ contribution, neglecting the effect of baryon
  couplings, shows the importance of the these contributions for the
  tremendous broadening of the $\rho$-spectral function as well as the
  yield in the very-low-mass range close to the two-muon
  threshold. Microscopically these contributions originate from Dalitz
  decays of various baryon resonances.}
\label{fig.1}
\end{figure}

All these microscopic processes have to be evaluated in a hot and dense
medium, which is at present available in detailed many-body
quantum-field theoretical calculations only for a medium in thermal
equilibrium, with some chemical off-equilibrium taken into account in
terms of effective baryon densities and pion- as well as kaon-chemical
potentials \cite{rw99b,gale-rapp99}. It has been quite well established
that the observed excess of low-mass dilepton production in heavy-ion
collisions as compared to the expectations from the ``hadronic
cocktail'' (i.e., from a simple scaling of the findings in pp collisions
by the number of nucleon collisions in the nuclear reaction) can be
understood by effective hadronic models that predict a tremendous
broadening of the light-vector mesons' spectral shape with small mass
shifts
\cite{vanHees:2006ng,vanHees:2007th,Dusling:2006yv,Ruppert:2007cr}.

On the other hand, one also has to describe the evolution of the hot and
dense medium in a reliable way. This is usually done either by employing
macroscopic descriptions, reaching from quite simple thermal-fireball
parametrizations (``blast-wave fits'') \cite{Rapp:2013nxa} to detailed
ideal or viscous hydrodynamics simulations
\cite{Teaney:2001av,Hirano:2002ds,Kolb:2003dz,Nonaka:2006yn,Dusling:2006yv,Vujanovic:2013jpa},
or microscopic transport simulations
\cite{Cassing:1997jz,Schenke:2005ry,Schenke:2006uh,Schenke:2007zz,Bratkovskaya:2008bf,Cassing:2009vt,Barz:2009yz,Linnyk:2011hz,Weil:2012ji,Weil:2012qh}.

In the latter approach it is challenging to fully implement the medium
modifications of the partonic and hadronic dilepton sources. In our
approach this problem is solved by mapping the kinetic description of
the fireball evolution via Monte-Carlo transport simulations to a
state close to local thermal equilibrium. In this work, an ensemble of
collision events simulated by the UrQMD model is used to calculate a
coarse-grained energy-momentum tensor and net-baryon-number current on a
space-time grid. Then, using the Eckart definition of the local rest
frame, a temperature and baryo-chemical potential can be extracted via
an equation of state (EoS), making use of a description of the
energy-momentum tensor in terms of an anisotropic-fluid ansatz
\cite{Florkowski:2010cf}, where the pressure anisotropy, particularly
relevant in the early stages of the fireball evolution, is taken into
account. For the EoS we match a parametrization of lattice-QCD results
\cite{He:2011zx} with a hadron-resonance gas EoS
\cite{Zschiesche:2002zr}, which match very well in the region of the
parton-hadron transition around the critical temperature of
$T_c \simeq 170 \; \MeV$.

In recent work \cite{Endres:2014zua,Endres:2015fna}, we have combined
this coarse-graining method for the bulk-medium evolution of the
strongly interacting fluid via the UrQMD transport model with
state-of-the art in-medium dilepton-production rates
\cite{rw99b,gale-rapp99} in order to evaluate the dilepton
invariant-mass and $p_T$ spectra in $158 A \, \GeV$ In+In collisions as
measured by the NA60 Collaboration at the CERN SPS and in
$1.76 A \, \GeV$ Ar+KCl and $1.23 A \, \GeV$ Au+Au collisions as
measured by the HADES collaboration at the GSI SIS.

\begin{figure}
\begin{minipage}{0.48\linewidth}
\includegraphics[width=1.0\columnwidth]{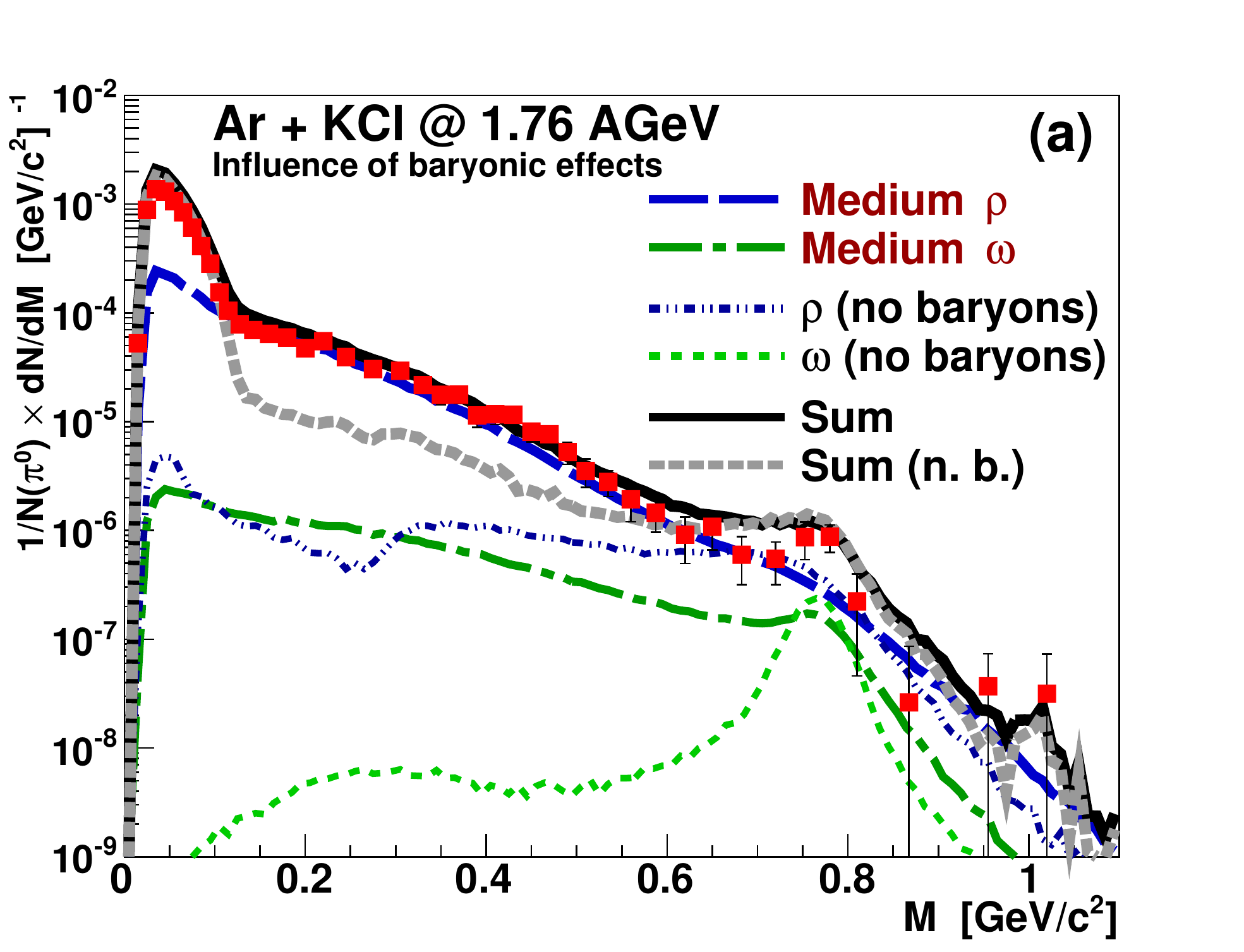}
\end{minipage}\hfill
\begin{minipage}{0.48\linewidth}
\includegraphics[width=1.0\columnwidth]{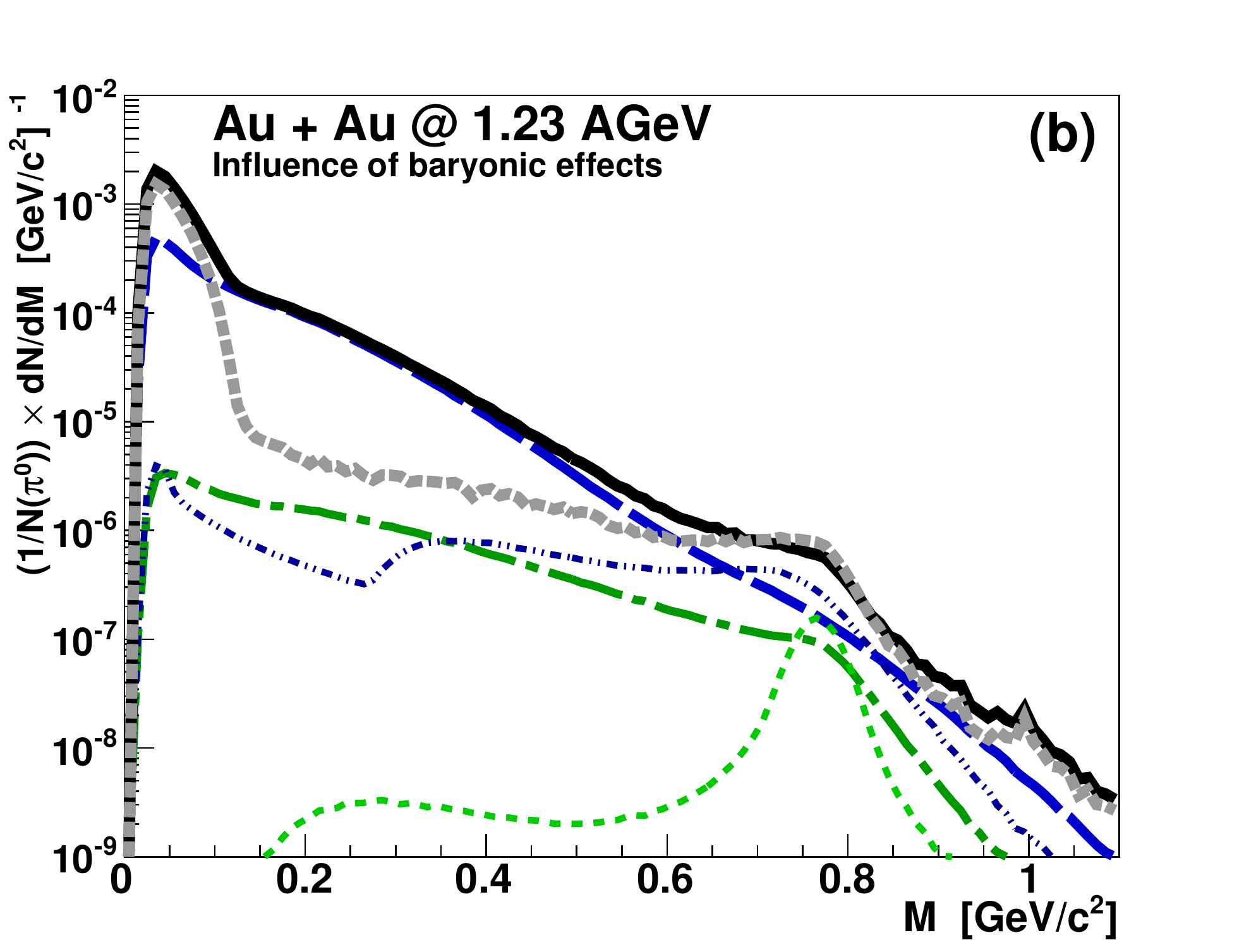}
\end{minipage}
\caption{(Color online) (a) Dielectron invariant-mass spectrum for
  Ar+KCl collisions at $E_{\mathrm{lab}}=1.76 A \, \GeV$ and (b) for
  Au+Au at $E_{\mathrm{lab}}=1.23 A \, \GeV$. The results are normalized
  to the average total number of $\pi^{0}$ per event and shown within
  the HADES acceptance. The results for Ar+KCl are compared to the data
  from the HADES Collaboration \cite{Agakishiev:2011vf}. The comparison
  of the dilepton spectra between using the full model and omitting the
  contributions of direct baryon couplings to the $\rho$ and $\omega$
  self-energies underlines the prevalence of the corresponding
  baryon-resonance Dalitz-decay processes processes for the in-medium
  modifications of these vector mesons, shifting substantial spectral
  strength to the low-invariant-mass region.}
\label{fig.2}
\end{figure}

As exemplified by the comparison of the so calculated dimuon
invariant-mass spectra in $158 A\, \GeV$ In+In Collisions with the NA60
data (Fig.\ \ref{fig.1}) and corresponding dielectron invariant-mass
spectra for $1.76 A\, \GeV$ Ar+KCl collisions with data from the HADES
collaboration (Fig.\ \ref{fig.2}, left panel), the model successfully
describes the data. In the right panel of Fig.\ \ref{fig.2} we
additionally provide our prediction for the $1.23 A\, \GeV$ Au+Au
collisions as also measured by the HADES collaboration.

Both figures underline the importance of the interaction of the light
vector mesons with baryons in the medium [in the present model
\cite{rw99} N(938), $\Delta(1232)$, N(1440), N(1520), $\Delta(1620)$,
$\Delta(1700)$, N(1720), $\Delta(1900)$, N(2000)]. As expected, these
processes are prevalent at the low GSI-SIS energy with high net-baryon
densities but also relevant at CERN-SPS energies and even at higher RHIC
energies. This is due to the fact that the strong interaction is CP
invariant and thus not the net-baryon density, $N_B-N_{\overline{B}}$,
but the total baryon-antibaryon density, $N_B+N_{\overline{B}}$, is
relevant for the in-medium modifications of the light-vector mesons'
spectral distributions.

In the near future we also plan to apply the model to dilepton
production at RHIC energies as well as photon production in heavy-ion
collisions.

\ack

We thank Ralf Rapp for providing the parameterization of the spectral
function and many fruitful discussions. This work was supported by BMBF,
HIC for FAIR and H-QM.

\section*{References}

\providecommand{\newblock}{}

\end{document}